\documentclass[conference,a4paper]{IEEEtran}

\usepackage{cite}
\usepackage{graphicx,color,epsfig,rotating}
\usepackage{amsfonts,amsmath,amssymb,bbm}
\usepackage{algorithm,algorithmic}
\usepackage{subfigure}
\usepackage{amsmath}
\usepackage{cite}
\usepackage{mdwtab}
\usepackage{placeins}
\usepackage{psfrag, graphicx}
\usepackage[latin1]{inputenc}
\usepackage{amssymb}
\usepackage{multirow}
\usepackage{mathrsfs}

\long\def\comment#1{}

\newfont{\bbb}{msbm10 scaled 700}

\newfont{\bb}{msbm10 scaled 1100}

\newcommand{\PP}{\mbox{\bb P}}

\newcommand{\FF}{\mbox{\bb F}}

\newcommand{\vv}{{\bf v}}

\newcommand{\Cm}{{\bf C}}

\newcommand{\Fm}{{\bf F}}

\newcommand{\Wm}{{\bf W}}
\newcommand{\Vm}{{\bf V}}

\newcommand{\Ec}{{\cal E}}
\newcommand{\Fc}{{\cal F}}

\newcommand{\Hc}{{\cal H}}

\newcommand{\Tc}{{\cal T}}
\newcommand{\Uc}{{\cal U}}

\newcommand{\Vc}{{\cal V}}

\newcommand{\omegav}{\hbox{\boldmath$\omega$}}

\newcommand{\fsf}{{\sf f}}

\newcommand{\transp}{{\sf T}}

\newcommand{\be}{\begin{equation}}
\newcommand{\ee}{\end{equation}}
\newcommand{\bea}{\begin{eqnarray}}
\newcommand{\eea}{\end{eqnarray}}

\def\fsf{ {\sf f}}

\newtheorem{defn}{Definition}
\newtheorem{theorem}{Theorem}

\begin{document}

\sloppy


\title{Caching and Coded Multicasting: Multiple Groupcast Index Coding}


 \author{
   \IEEEauthorblockN{
     Mingyue Ji\IEEEauthorrefmark{1},
     Antonia M. Tulino\IEEEauthorrefmark{2},
     Jaime Llorca\IEEEauthorrefmark{2} and 
     Giuseppe Caire\IEEEauthorrefmark{1}\IEEEauthorrefmark{3}  
    }
   \IEEEauthorblockA{
     \IEEEauthorrefmark{1}
     EE Department, University of Southern California.  Email: \{mingyuej, caire\}@usc.edu}
   \IEEEauthorblockA{
     \IEEEauthorrefmark{2} Alcatel Lucent, Bell Labs,Holmdel, NJ, USA. 
     Email: \{a.tulino, jaime.llorca\}@alcatel-lucent.com}
     \IEEEauthorblockA{
     \IEEEauthorrefmark{3} EECS Department, Technische Universit\"at Berlin. Email: caire@tu-berlin.de}
 }

\maketitle

\begin{abstract}
The capacity of caching networks has received considerable attention in the past few years. 
A particularly studied setting is the case of a single server (e.g., a base station) and multiple users, each of which caches segments of 
files in a finite library. Each user requests one (whole) file in the library and the server sends a common coded multicast message to 
satisfy all users at once. The problem consists of finding the smallest possible codeword length to satisfy such requests. 
In this paper we consider the generalization to the case where each user places $L \geq 1$ requests. 
The obvious naive scheme consists of applying $L$ times the order-optimal scheme 
for a single request,  obtaining a linear in $L$ scaling of the multicast codeword length. 
We propose a new achievable scheme based on multiple groupcast index coding that achieves a significant gain over the naive scheme. 
Furthermore, through an information theoretic converse we find that the proposed scheme is approximately optimal
within a constant factor of (at most) $18$. 
\end{abstract}

\section{Introduction}
\label{section: intro}

Wireless and mobile data traffic has grown 
dramatically in the past few years and is expected to increase at an even faster pace in the near future, 
mainly pushed by on-demand video streaming \cite{cisco13}. Such type of traffic is characterized by {\em asynchronous content reuse} \cite{6495773}, i.e., 
the users demands concentrate on a relatively small library of files (e.g., 1000 titles of TV shows and movies), 
but the streaming sessions happen at arbitrary times  such that {\em naive multicasting} of the same file (e.g., by exploiting the inherent broadcast 
property of the wireless channel) yields no significant gain.   A classical and effective approach to leverage such asynchronous content reuse 
is {\em caching} at the user devices.  A significant amount of recent work has shown that caching at the wireless edge 
yields very effective ways to trade-off expensive wireless bandwidth for cheap and widely available storage memory in the user devices, 
which is the fastest growing and yet untapped network resource in today's wireless networks \cite{llorcatulino132 , llorcatulino14, ji2013throughput, ji2013fundamental, ji2013wireless, ji2014order, molisch2014caching, golrezaei2012device, maddah2012fundamental,maddah2013decentralized,gitzenis2012asymptotic}. 

In \cite{llorcatulino132 , llorcatulino14}, Llorca {\it et al.} formulated a general framework of the caching problem 
in an arbitrary network as an optimization problem and showed the effectiveness of coded delivery via simulations. They further showed the NP-hardness of the problem via an equivalence to network coding in a caching-demand augmented graph. 
However, due to the generality of the problem, except for a few cases it is unlikely to obtain the optimality of the achievable schemes and the gap between the achievable rate and a converse bound. 

In \cite{maddah2012fundamental , maddah2013decentralized}, Maddah-Ali and Niesen 
considered a network formed by a single server (a base station) and $n$ users, where 
each user place an arbitrary demand to a file in the library and the server sends a single multicast codeword that is received by all users and satisfies at once
all the demands. An achievable scheme based on a combinatorial cache construction (caching phase) and linear coding for the multicast transmission 
(delivery phase) is proposed in  \cite{maddah2012fundamental} and, through a cut-set information theoretic bound, it is shown that such scheme is order-optimal in the case of worst-case demands, i.e., in a min-max sense (min codeword length, max over the user demands). 
This result is extended to the case of random decentralized caching phase in \cite{maddah2013decentralized}. 

Consistently with \cite{maddah2012fundamental , maddah2013decentralized}, in this paper
we refer to the {\em transmission rate} as the length (expressed in equivalent file units) of the multicast codeword. 
The order-optimal transmission rate scaling found in \cite{maddah2012fundamental , maddah2013decentralized} is given by 
$\Theta\left(\min\left\{\frac{m}{M}, m, n\right\}\right)$. Notice that in the interesting regime where $n$ and $m$ are large, $M$ is fixed
and $nM \gg m$, this yields a multiplicative factor of $M$ in the per-user throughput (in bits per unit time). 
In passing, we notice here that an alternative caching and delivery approach was proposed 
in \cite{ji2013throughput, ji2013fundamental}, where the communication is also one-hop but takes place between the user nodes
(Device-to-Device or ``D2D'' communications). Both in the random decentralized caching case  \cite{ji2013throughput} 
and in the combinatorial caching and coded delivery case \cite{ji2013fundamental}, the same rate scaling law is shown to be achievable 
and approximately optimal for the one-hop D2D network also. In light of these results, it appears that the problem of caching at the user nodes
(either in a combinatorial/centralized or random decentralized manner) and delivery (either coded multicast from a single server 
\cite{maddah2012fundamental , maddah2013decentralized}, or uncoded with individual peer-to-peer transmissions \cite{ji2013throughput}, 
or coded in multiple peer-to-multicast transmissions \cite{ji2013fundamental}) is well-understood. 

In this work, we consider the same setting of \cite{maddah2012fundamental} (one server, $n$ users, one-hop multicast transmission from the server to the users)
where users make multiple requests instead of a single request. This scenario may be motivated by a {\em FemtoCaching} network \cite{6495773} 
formed by $n$ small-cell base stations receiving data from a controlling ``macro'' base station via the cellular downlink. 
Each small-cell base station has a local cache of $M$ file units and serves $L$ users through its own local high-rate downlink. 
Hence, each small-cell base station (which is identified as a ``user'' in our network model) makes $L$ requests to the macro base station at once. 
{A related work is presented in \cite{maddah2013decentralized}, 
under the constraint that  each user's storage capacity scales linearly with the number of the requests per user. This is referred to as \emph{Shared Caches}. In addition, \cite{maddah2013decentralized} only studied the case that $m > nL$ or every user requests distinct files for the worst-case demands. This can be easily violated if the library is considered as the most popular files due to asynchronous content reuse.}
Another related relevant work is presented in \cite{6620404}, where an exact solution of the 3-user index coding problem for assigned side information 
and multiple requests is given. We study the fundamental limits of this type of network for
the general case of $n$ users, $m$ possible messages, storage capacity  $M$ and $L$ requests per user without any constraint, where in contrast to the general index coding problem, 
the side information (i.e., the cache content) at each user is designed as part of the ``code'' rather than given a priori. 

Our contribution is two-fold. First, by using the {\em same}  combinatorial caching phase of  \cite{maddah2012fundamental}, we generalize the 
coded delivery phase by using the directed (fractional) local chromatic number proposed in \cite{shanmugam2013local} to the case of multiple groupcasting, 
where {\em multiple} means that each user makes $L \geq 1$ requests and {\em groupcasting} means that 
one file or packet (see later) may be requested by several users. 
We show that order gains can be obtained by using the proposed scheme compared to the naive approach of using $L$ times the delivery phase of \cite{maddah2012fundamental}. Second, we present an information theoretical lower bound of the rate, and show that the proposed scheme
meets this lower bound within a constant factor of at most $18$. 

\section{Network Model and Problem Formulation}
\label{section: network model}

We consider a network with $n$ user nodes $\mathcal U = \{1, \cdots, n\}$ connected through a single bottleneck link 
to a server. The server has access to the whole content library $\mathcal{F} = \{1, \cdots, m\}$ containing $m$ files of same size of $B$ bits.
Each user node has a cache of size $M$ files (i.e., $MB$ bits).  
The bottleneck link is a shared deterministic channel that transmits one file per unit time, such that all the users can decode the same multicast codeword. 
At each time unit (slot), each user requests an arbitrary set of $L$ files in $\mathcal F$. Such requests form a matrix $\Fm$ of size $L \times n$
with columns  $\fsf_u = (f_{u, 1}, f_{u, 2} \ldots, f_{u, L})^{\rm T}$ corresponding to the requests of each user $u \in \Uc$. 
The caching problem includes two distinct operations: the caching phase and the delivery phase. The caching phase (cache formation) is done a priori, 
as a function of the files in the library, but does not depend on the request matrix $\Fm$. Then, at each time slot, given the current request matrix $\Fm$,  
the server forms a multicast codeword and transmits it over the bottleneck link such that all users can decode their requested files. 
Formally, we have: 

\begin{defn}
{\bf (Caching Phase)} The caching phase is a map of the file library $\Fc$ onto the user caches. Without loss of generality, 
we represent files as vectors over the binary field $\FF_2$. For each $u \in \Uc$, let $\phi_u: \FF_2^{mB} \rightarrow \FF_2^{MB}$ denote the caching function of
user $u$. Then, the cache content of user $u$ is given by $Z_u \triangleq \phi_u(W_f : f = 1, \cdots, m)$, where $W_f \in \FF_2^{B}$ denotes the $f$-th file in the library. 
\hfill $\lozenge$
\end{defn}

\begin{defn}
{\bf (Delivery Phase)} 
A delivery code of rate $R(M,L)$ is defined by an encoding function $\psi : \FF_2^{mB} \times \Fc^{L \times n} \rightarrow \FF_2^{R(M,L)B}$ 
that generates the codeword $X_{{\bf F}}  = \psi(\{W_f : f = 1, \cdots, m\}, {\bf F})$ transmitted by the server to the users, 
and decoding functions $\lambda_u : \FF_2^{R(M,L)B} \times \FF_2^{MB} \times \Fc^{L \times n} \rightarrow \FF_2^{LB}$ such that 
each user $u \in \Uc$ decodes its requested files as
$(\hat{W}_{u, f_{u,1}}, \ldots, \hat{W}_{u, f_{u,L}}) = \lambda_u( X_{{\bf F}}, Z_u, {\bf F})$.
\hfill $\lozenge$
\end{defn}

The case of arbitrary demands can be formulated as a {\em compound channel}, where the delivery phase is designed in order to minimize the 
rate for the worst-case user demand. The relevant worst-case error probability is defined as
\begin{equation}
P_e = \max_{{\bf F}} \max_{u \in \Uc} \max_{\ell = 1, \ldots, L} \; \PP\left( \hat{W}_{u, f_{u,\ell}}  \neq W_{f_{u,\ell}} \right).
\end{equation}
The cache-rate pair $(M, R(M, L))$ is achievable if there exist a sequence of codes $\{\phi , \psi, \{ \lambda_u : u \in \Uc\} \}$ for increasing file size 
$B$ such that 
$\lim_{B \rightarrow \infty} P_e = 0$. In this context, the system capacity $R^*(M, L)$ (best possible achievable rate) is given by the 
infimum of all $R(M,L)$ such that $(M, R(M, L))$ is achievable. In the rest of the paper, proofs are omitted for the sake of space limitation, and are provided in \cite{mingyue20142}.

\section{Achievable Scheme}
\label{sec: Achievable Scheme}

In this section, we present the proposed achievable scheme. 
Since worst-case demands are considered, we simply use the same {\em packetized caching function} defined in \cite{maddah2012fundamental}, 
which is optimal (within a fixed factor) for the case $L = 1$. For the sake of completeness, we describe the caching scheme in the following. 
Let $t = \frac{Mn}{m}$ be a positive integer, then we consider the set $\mathcal{P}$ of all subsets $\mathcal{T}$ (combinations) of distinct users of size $t$. 
Each file is divided into ${n \choose t}$ packets. For each file, we use $\mathcal{T} \in \mathcal{P}$ to label all the file packets. 
Then, user $u$ will cache the packets whose label $\Tc$ contains $u$. 
We denote by $\mathcal{S}$ the set of all implicitly requested packets.

Based on this caching scheme, we design a delivery scheme based on linear index coding 
(i.e., $\phi$ is  linear function over an extension field of $\FF_2$). In particular, we focus on encoding functions of the following form: for a 
request matrix $\Fm$, the multicast codeword is given by
\be  \label{VV}
X_{\bf F} = \sum_{s \in \mathcal{S}} \omega_s \vv_s = \Vm \omegav,
\ee
where 
$\omega_s$ is the binary vector corresponding to packet $s$, represented as a (scalar) symbol of
the extension field $\FF_\kappa$ with $\kappa = {2^{B/{n \choose t}}}$,  the $\nu$-dimensional vector $\vv_s \in \FF_\kappa^\nu$ is the coding vector of packet $s$ and where
we let $\Vm = [\vv_1, \cdots. \vv_{|\mathcal{S}|}]$ and $\omegav= [\omega_1 ,\cdots, \omega_{|\mathcal{S}|}]^{\transp}$. 
The number of rows $\nu$ of $\Vm$ yields the number of packet transmissions. Hence, the 
coding rate is given by $R(M,L) = \nu/{n \choose t}$ file units.  

In order to design the coding matrix $\Vm$ for multiple group-casting, we shall use the method based on directed local chromatic number 
introduced in \cite{shanmugam2013local}. The definition of directed local chromatic number is given as follows: 
\begin{defn} ({\bf Directed Local Chromatic Number (I)})
The directed local chromatic number of a directed graph $\mathcal{H}^d$ is defined as: 
\begin{align}
& \chi_l(\mathcal H^d) = \min_{c \in \mathcal{C}} \max_{v\in\mathcal{V}}|c(\mathcal{N}^+(v))|
\end{align}
where $\mathcal{C}$ denotes the set of all vertex-colorings of $\mathcal{H}$, the undirected version of $\mathcal H^d$ obtained by ignoring the directions of edges in $\mathcal{H}^d$, 
$\mathcal{V}$ denotes the vertices of $\mathcal{H}^d$, $\mathcal{N}^+(v)$ is the closed out-neighborhood of 
vertex $v$,\footnote{Closed out-neighborhood of vertex $v$ includes vertex $v$ and all the connected vertices 
via out-going edges of $v$.} 
and $c(\mathcal{N}^+(v))$ is the total number of colors in $\mathcal{N}^+(v)$ for the given coloring $v$. 
\hfill $\lozenge$
\end{defn}

An equivalent definition of the directed local chromatic number
in terms of an optimization problem is given by:

\begin{defn} ({\bf Directed Local Chromatic Number (II)})
Let $\mathcal I$ denote the set of all independent sets of $\mathcal H$, the directed local chromatic number of 
$\mathcal H^d$ is given by: 
\begin{align}\label{local}
\text{minimize} &\quad k \notag\\
\text{subject to} 
& \quad \sum_{I:\mathcal N^+(v) \bigcap I \neq \emptyset} x_I \leq k, \quad \forall v\in\mathcal V \\
& \quad \sum_{I:v\in I} x_I \geq 1, \quad \forall v\in\mathcal V \\
& \quad x_I \in\{0,1\}, \quad \forall I\in\mathcal I \label{eq: integer}
\end{align}
\hfill $\lozenge$
\end{defn}

Also, we have:

\begin{defn} ({\bf Directed Fractional  Local Chromatic Number}) \label{def:local}
The directed fractional chromatic number is given by (\ref{local}) when relaxing (\ref{eq: integer}) to $x_I \in [0,1]$. 
\hfill $\lozenge$
\end{defn}

In the following, we describe the proposed scheme when $t$ is a positive integer. 
When $t$ is not an integer, we can simply use the resource sharing scheme for caching 
proposed in \cite{maddah2012fundamental} and achieve convex combinations of the cases where $t$ is an integer. 

In order to find the coding matrix $\Vm$ we proceed in three steps \cite{shanmugam2013local}: 
1) constructing the directed conflict graph $\mathcal{H}^d$; 
2) computing the directed local chromatic number $\chi_l(\mathcal H^d)$ and the corresponding vertex-coloring $c$; 
3) constructing $\Vm$ by using the columns of the generator matrix of a $(|c|,\chi_l(\mathcal H^d))$-MDS code.\footnote{According to the classical coding theory notation,  an $(\rho,\nu,d)$-MDS code is a code with length $\rho$, dimension $\nu$, and minimum Hamming distance $d = \rho-\nu + 1$. 
An MDS code is able to correct up to $\rho-\nu$ erasures. This implies that a linear MDS code has generator matrix of dimensions
$\nu \times \rho$ such that any subset of $\ell \leq \nu$ columns form a $\nu \times \ell$ submatrix of rank $\ell$.} 
The detailed delivery scheme is described in the following.

\noindent
1) Construction of the directed conflict graph $\mathcal{H}^d$.
\begin{itemize}
\item Consider each packet requested by a user as a {\em distinct} vertex in $\mathcal{H}^d$. 
This means that if  a certain packet is requested $N$ times, for some $N$, because it appears in the requests of different users, 
it corresponds to $N$ different vertices in $\Hc^d$. Hence, each vertex of $\mathcal H_{\Cm, \Wm} = (\Vc, \Ec)$ is uniquely  identified by the pair 
$v = \{ \rho(v),\mu(v)\}$  where $\rho(v)$ indicates the  \mbox{packet identity} associated to the vertex and 
$\mu(v)$ represents the \mbox{user requesting it}.
Since each user caches 
${n-1 \choose t - 1}$ packets of each file, and each file is partitioned into ${n \choose t}$ packets, 
the number of requested packets of each file for each user is ${n \choose t} - {n-1 \choose t - 1} = {n - 1 \choose t}$ (by Pascal's triangle). 
Hence, the total number of files requested $Ln$ corresponds to $|\mathcal{V}| = Ln  {n - 1 \choose t}$ vertices in $\mathcal{H}^d$.
\item For any pair of vertices $v_1$, $v_2$, we say that vertex (packet) $v_1$ interferes with vertex $v_2$ 
if the packet associated to the vertex $v_1$, $\rho(v_1)$,  is not  in the cache of the user associated to vertex $v_2$, namely, $\mu(v_2)$. 
In addition, $\rho(v_1)$ and $\rho(v_2)$ do not represent the same packet. Then, 
draw a directed edge from vertex $v_2$ to vertex $v_1$ if $v_1$ interferes with $v_2$.
\end{itemize}

\noindent
2) Color the directed conflict graph $\mathcal{H}^d$ according to the coloring which gives the directed local chromatic number 
$\chi_l(\mathcal{H}^d)$. The total number of colors needed to give the directed local chromatic number is denoted by $|c|$. 

\noindent
3) Construct $\Vm'$ be the generator matrix of an MDS code with parameters $(|c|,\chi_l(\mathcal{H}^d))$ over
$\FF_\kappa$. Notice that it is well-known that for sufficiently large $\kappa$ such
MDS code exists. Since we are interested in $B \rightarrow \infty$ and $\kappa = 2^{B/{n \choose t}}$, for sufficiently large $B$ this code exists. 

\noindent
4) Allocate the same coding vector to all the vertices (packets) with the same color in $c$. Then, $\Vm$ is obtained from the MDS generator matrix $\Vm'$ 
by replication of the columns, such that each column in $\Vm$ is replicated for all vertices with the same corresponding color in $c$. 
Eventually, all packets are encoded using the linear operation in (\ref{VV}). 

The above constructive coding scheme proves the following (achievable) upper bound on the optimal coding length:

\begin{theorem}
\label{theorem: 1}
Under worst-case demand, the optimal coding length for the multiple group cast problem  with integer $t = nM/m$ satisfies
\be
R^*(M, L) \leq R^{\rm LC}(M, L) =  \max_{\Fm} \; \frac{\chi_l(\mathcal{H}^d)}{{n \choose t}}.
\ee
where the upper bound is achieved by the caching and linear coded delivery scheme seen above. 
\hfill  $\square$
\end{theorem}

We can also consider a delivery scheme based on the directed fractional local chromatic number, as given in Definition \ref{def:local},
that achieves a generally better performance. 
For the optimality of index coding based on local chromatic number, where only the index coding delivery phase 
is considered for assigned node side information (i.e., without optimizing caching functions), it can be shown that the the gap 
between the proposed index code and the converse is bounded by the integrality gap of a certain linear programming (see \cite{yu2013duality}). 

\section{Performance Analysis}
\label{sec: Performance Analysis}

To show the effectiveness of the proposed coded caching and delivery scheme, we first consider the two special 
points of the achievable rate versus $L$ corresponding to $L=1$, considered in \cite{maddah2012fundamental}, 
and $L = m$, i.e., when every user requests the whole library. 

\begin{theorem}
\label{theorem: one request}
When each user makes only one request ($L=1$), then the achievable rate of the proposed scheme satisfies
\be
\label{eq: one request}
R^{\rm LC}(M, 1) \leq R^{\rm MN}(M),
\ee
where $R^{\rm MN}(M)$ is the convex lower envelope of 
$\min\left\{n\left(1-\frac{M}{m}\right)\frac{1}{1+\frac{Mn}{m}}, m-M\right\}$, achieved by the scheme of \cite{maddah2012fundamental}.
\hfill  $\square$
\end{theorem}

\begin{theorem}
\label{theorem: 3}
When each user requests the whole library ($L=m$), then the achievable rate of the proposed scheme satisfies 
\be
\label{eq: m request}
R^{\rm LC}(M,m) \leq m - M.
\ee
\hfill  $\square$
\end{theorem}

The rate $m - M$ can also be achieved, with high probability for a large field size, 
by using random linear coding. Moreover, it can be shown that this is information theoretically 
optimal.

Theorems \ref{theorem: one request} and \ref{theorem: 3} show that in the extreme regimes of $L$ the proposed scheme performs at least as good 
as the state of the art scheme in literature. 
A general upper bound of the achievable rate is given by:
\begin{theorem}
The achievable rate of the proposed scheme satisfies
\be
\label{eq: upper bound FLC}
R^{\rm LC}(M,L) \leq R^{\rm LC}_{\rm ub}(M,L)
\ee
where $R^{\rm LC}_{\rm ub}(M,L)$ is the convex envelope (with respect to $M$) of 
\[ \min\left\{ Ln\left(1-\frac{M}{m}\right)\frac{1}{1+\frac{Mn}{m}}, m-M \right \}. \]
\hfill  $\square$
\end{theorem}

\begin{figure}[ht]
\centering
\subfigure[]{
\centering
\includegraphics[width=4.1cm]{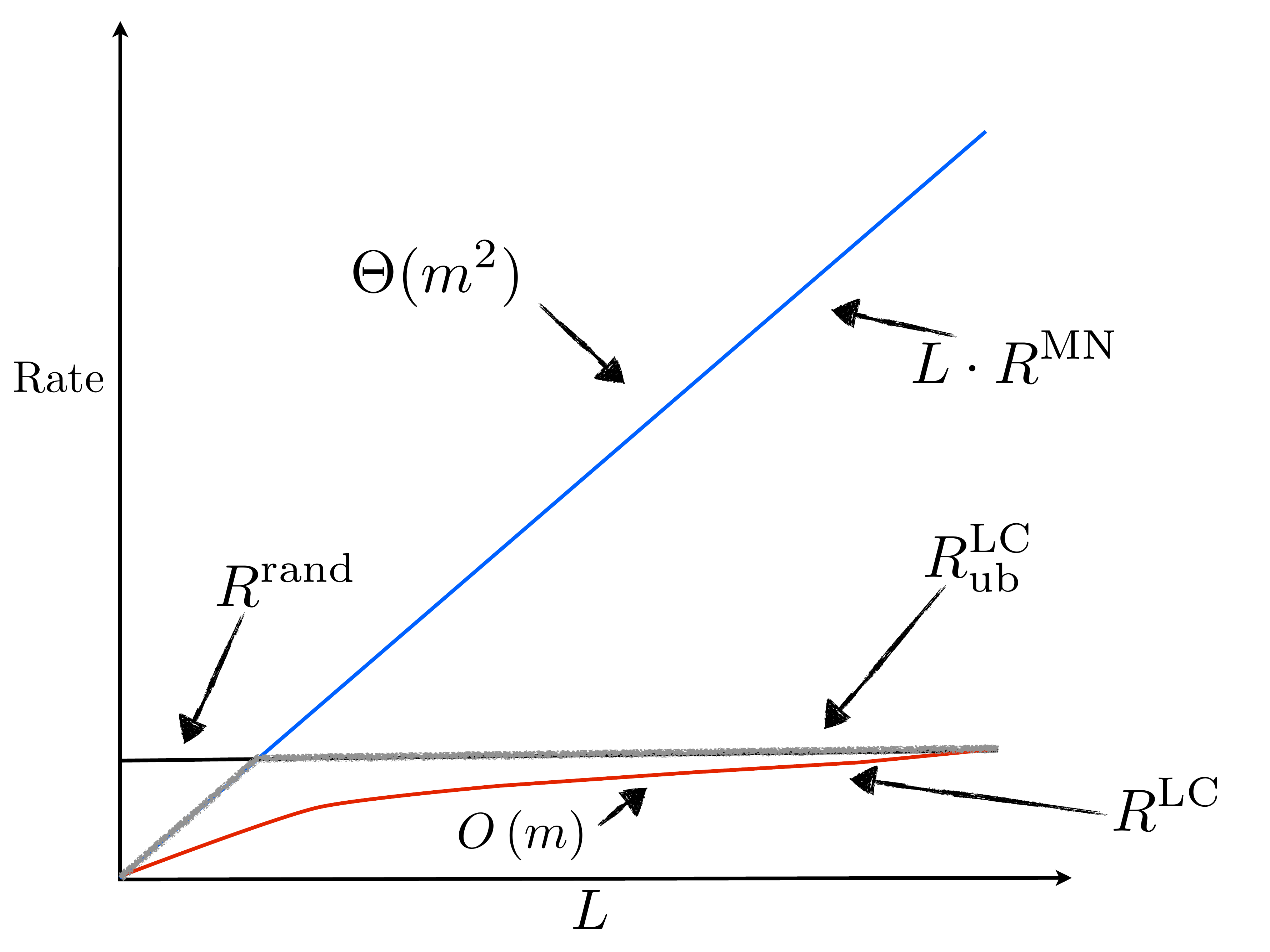} 
\label{fig: Performance_1}
}
\subfigure[]{
\centering
\includegraphics[width=4.1cm]{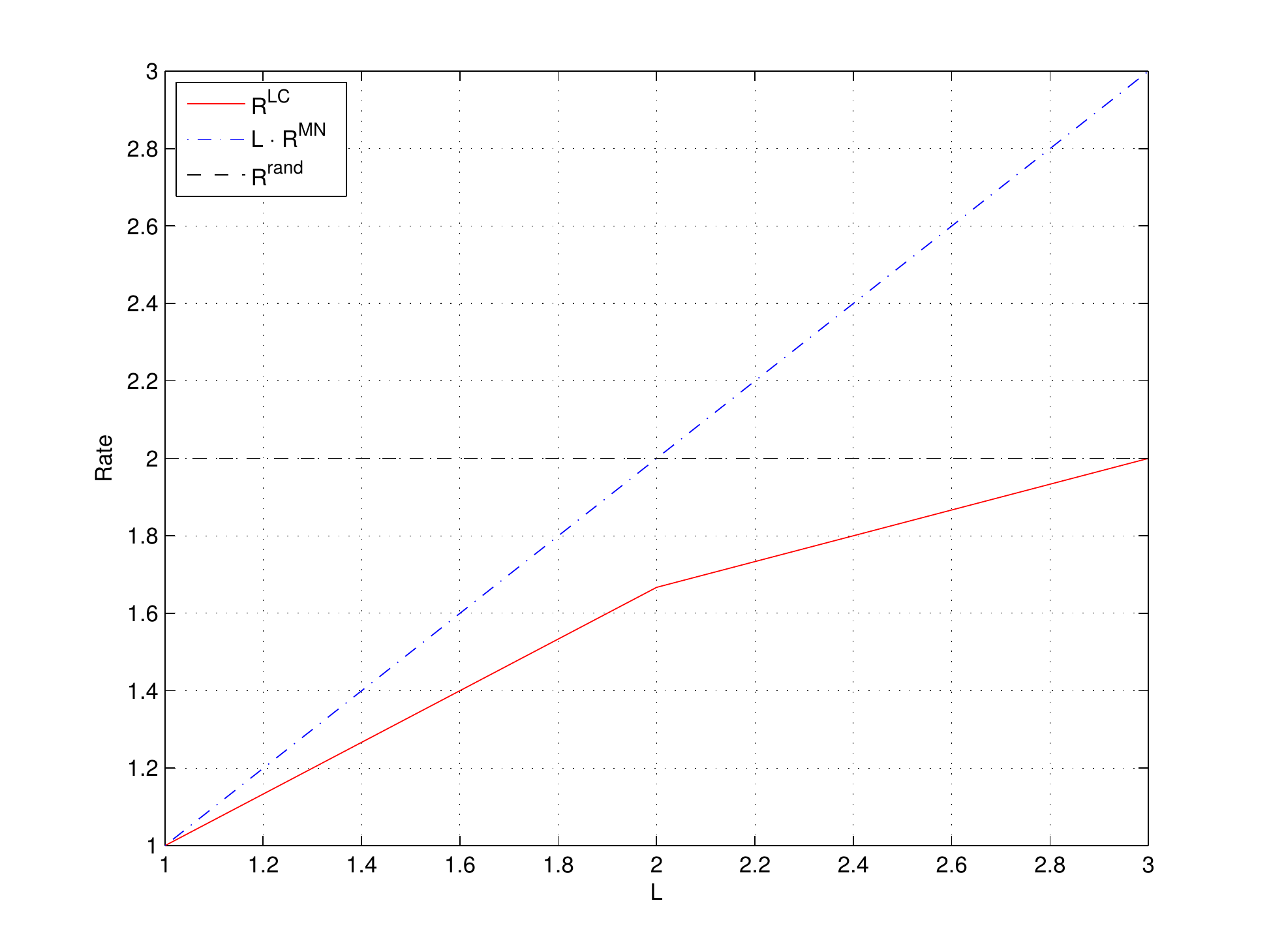} 
\label{fig: Performance_2}
}
\caption{(a).
A qualitative illustration of the different delivery schemes ($M$ is assumed to be not very small). $L \cdot R^{\rm MN}$ (blue curve) represents rate by using the scheme in \cite{maddah2012fundamental} $L$ times. $R^{\rm rand}$ (black curve) is the rate by random linear coding. $R^{\rm LC}_{\rm ub}$ (grey curve) is an upper bound of the achievable rate of the proposed scheme. $R^{\rm LC}$ (red curve) represents the rate by the proposed scheme based on directed local chromatic number. (b).~An example of the rate by the proposed scheme. In this example, $n=m=3$ and $M=1$. In this figure, 
all the symbols have the same meanings as in Fig. \ref{fig: Performance_1}. }
\end{figure}

The qualitative performance of the proposed scheme is shown in Fig. \ref{fig: Performance_1}, where 
we call the scheme that repeats the delivery scheme designed for one request (in \cite{maddah2012fundamental}) $L$ times 
as \emph{direct scheme}. To measure the performance of the proposed scheme quantitatively, we let $L = \alpha m$, 
where $\alpha \in (0,1)$. Then, let $M$ be a constant, as $m \rightarrow \infty$, we can see that the rate of the direct scheme is $L \cdot R^{\rm MN}$, 
which scales as $\Theta(m^2)$. While by using (\ref{eq: upper bound FLC}), the rate of the proposed scheme scales at most $O(m)$. 
Thus, we can see the proposed scheme can have an order gain compared to the direct scheme.

In order to appreciate the gain achieved by the proposed scheme over the direct scheme for fixed parameters, 
we consider the case of $n=m=3$ and $M=1$. 
In this case, though,  the requests can be arbitrary. 
Fig. \ref{fig: Performance_2} shows the worst-case (over the requests) coding rate versus $L$. We can observe 
that, even for small $n$, $m$ and $M$, the gain achieved by the proposed scheme with respect to the direct scheme is fairly large. 
For example, for $L=2$, the proposed scheme requires $\frac{5}{3}$ file units to satisfy any request, while the directed scheme or 
random linear coding require $2$ file units.

%

\section{Converse and Optimality}
\label{sec: Converse}

To show the optimality (up to a constant factor) of the proposed scheme, we prove an information theoretic lower bound on 
the rate by using the cut-set bound technique: 
\begin{theorem}
\label{theorem: lower bound}
Under worst-case demand, the optimal coding length for the multiple group cast problem satisfies
\begin{eqnarray}
\label{eq: lower bound 1}
&& R^*(M,L) \geq R^{\rm lb}(M, L) = \notag\\
&& \max\left\{\max_{s = 1, \cdots, \min\{ \lfloor\frac{m}{L}\rfloor, n\}}\left(Ls - \frac{sM}{\left\lfloor \frac{\lfloor\frac{m}{L}\rfloor}{s} \right\rfloor}\right), 
 \frac{m-M}{\lceil\frac{m}{L} \rceil} \right\}. \notag 
\end{eqnarray}
\hfill  $\square$
\end{theorem}
Using Theorem \ref{eq: lower bound 1}, we show that:
\begin{theorem}
\label{theorem: gap}
The multiplicative gap between the upper and lower bounds satisfies
\begin{eqnarray}
\frac{R^{\rm LC}(M,L)}{R^{\rm lb}(M,L)} \leq 18.
\end{eqnarray}
\hfill  $\square$
\end{theorem}

From simulation, we observed that this multiplicative gap is generally smaller than $5$.
For example, when $m=n=100$ and $M=20$, 
the we found a gap always less than $3.88$. 

\section{Discussions}
\label{sec: Discussion and Conclusion}

For the index coding problem, since the side information is assigned,
it is generally difficult to find a constant multiplicative gap between the achievable rate and the converse. 
For example, in \cite{haviv2012linear}, the authors consider a random unicast index coding problem, where each user 
can cache $\beta m$ files uniformly and $\beta \in (0,1)$. It is assumed that $m=n$ and all the users request different files. 
In this case, the best {\em multiplicative} gap between the achievable rate and the converse lower bound is $\Theta\left(\frac{\sqrt{n}}{\log n}\right)$, 
which goes to infinity as $n \rightarrow \infty$. However, for the caching problem, due to the fact that we can design the caching functions, 
it is possible to find coding schemes with a bounded multiplicative gap for the worst-case demands. 
In the case treated in this paper, for example, any scheme (possibly non-linear) would provide at most a factor of $18 \times$ gain with respect to the proposed
linear scheme.   

We notice from Section \ref{sec: Performance Analysis} that, to achieve the order gain compared with the direct scheme, 
we need to code over all the $L$ requested files simultaneously, in contrast with the repeated application of the scheme in 
\cite{maddah2012fundamental} (direct scheme), where the user can decode instantaneously. 
Therefore, the proposed scheme may be useful in the FemtoCaching application, where each user in our system corresponds to a local server (a small-cell base station) serving $L$ requests to its own local users on a much faster local connection. In this case, there is no natural ordering of the $L$ requests such that
there is no interest to decode them instantaneously, however based on the applications we can choose which requests combine together. 

\bibliographystyle{IEEEbib}
\bibliography{references}

\end{document}